\documentclass[twocolumn]{aastex63}
 
\newcommand{\msun}{$M_{\sun}$}

\newcommand{\msunyr}{\msun\,yr$^{-1}$}
\newcommand{\halpha}{H$\alpha$}

\newcommand{\mdot}{$\dot{M}$}
\newcommand{\ri}{R$_{\rm i}$}
\newcommand{\rw}{W$_{\rm r}$}
\newcommand{\tmax}{T$_{\rm max}$}

\newcommand{\kms}{km~s$^{-1}$}

\newcommand{\neii}{[\ion{Ne}{2}]}
\newcommand{\neiii}{[\ion{Ne}{3}]}

\usepackage{graphicx}
\usepackage[caption=false]{subfig}
\usepackage{xcolor}
\usepackage{url}
\usepackage{hyperref}

\newcounter{column_number}
\shortauthors{Espaillat et al.}
\shorttitle{Neon Line Variability}

\begin{document}

\title{{\it JWST} Detects Neon Line Variability in a Protoplanetary Disk}



\correspondingauthor{Catherine C. Espaillat}
\email{cce@bu.edu}

\author[0000-0001-9227-5949]{C. C. Espaillat}
\affil{Institute for Astrophysical Research, Department of Astronomy, Boston University, 725 Commonwealth Avenue, Boston, MA 02215, USA}

\author[0000-0003-4507-1710]{T. Thanathibodee}
\affil{Institute for Astrophysical Research, Department of Astronomy, Boston University, 725 Commonwealth Avenue, Boston, MA 02215, USA}

\author[0000-0001-9301-6252]{C. V. Pittman}
\affil{Institute for Astrophysical Research, Department of Astronomy, Boston University, 725 Commonwealth Avenue, Boston, MA 02215, USA}

\author[0000-0002-0377-1316]{J.~A. Sturm} 
\affil{Leiden Observatory, Leiden University, PO Box 9513, NL–2300 RA Leiden, The Netherlands}

\author[0000-0003-1878-327X]{M.~K. McClure} 
\affil{Leiden Observatory, Leiden University, PO Box 9513, NL–2300 RA Leiden, The Netherlands}

\author[0000-0002-3950-5386]{N. Calvet}
\affil{Department of Astronomy, University of Michigan, 1085 South University Avenue, Ann Arbor, MI 48109, USA} 

\author[0000-0001-7796-1756]{F.~M. Walter}
\affil{Department of Physics and Astronomy, Stony Brook University, Stony Brook, NY 11794-3800, USA}

\author[0000-0002-1650-3740]{R. Franco-Hern{\'a}ndez}
\affil{Instituto de Astronom{\'i}a y Meteorolog{\'i}a, Universidad de Guadalajara, Avenida Vallarta No. 2602, Col. Arcos Vallarta Sur, CP 44130, Guadalajara, Jalisco, Mexico} 

\author[0000-0002-5943-1222]{J. Muzerolle Page}
\affil{Space Telescope Science Institute, 3700 San Martin Drive, Baltimore, MD 21218, USA}

\begin{abstract} 

We report the first detection of variability in the mid-infrared neon line emission of a protoplanetary disk by comparing a {\it JWST} MIRI MRS spectrum of SZ Cha taken in 2023 with a {\it Spitzer} IRS SH spectrum of this object from 2008. We measure the [\ion{Ne}{3}]-to-[\ion{Ne}{2}] line flux ratio, which is a diagnostic of the high-energy radiation field, to distinguish between the dominance of EUV- or X-ray--driven disk photoevaporation. We find that the [\ion{Ne}{3}]-to-[\ion{Ne}{2}] line flux ratio changes significantly from $\sim1.4$ in 2008 to $\sim0.2$ in 2023. This points to a switch from EUV-dominated to X-ray--dominated photoevaporation of the disk. We present contemporaneous ground-based optical spectra of the {\halpha} emission line that show the presence of a strong wind in 2023. We propose that this strong wind prevents EUV radiation from reaching the disk surface while the X-rays permeate the wind and irradiate the disk. We speculate that at the time of the {\it Spitzer} observations, the wind was suppressed and EUV radiation reached the disk. These observations confirm that the MIR neon emission lines are sensitive to changes in high-energy radiation reaching the disk surface. This highlights the [\ion{Ne}{3}]-to-[\ion{Ne}{2}] line flux ratio as a tool to gauge the efficiency of disk photoevaporation in order to provide constraints on the planet-formation timescale. However, multiwavelength observations are crucial to interpret the observations and properly consider the star-disk connection.

\end{abstract}


\section{Introduction} \label{intro}

The protoplanetary disk lifetime is influenced by the rate at which gas is eroded by photoevaporative winds created by high-energy radiation from the central star \citep[see review by][]{pascucci22}. However, photoevaporation models predict significantly different mass-loss rates depending on whether the photoevaporation is driven by the extreme ultraviolet (EUV) or the X-ray radiation field. These different mass-loss rates lead to different timescales for disk clearing and hence the timescales by which giant planets must form. Models of EUV photoevaporation predict low mass-loss rates of $\sim$10$^{-10}$ {\msunyr} \citep{font04, alexander06} while X-ray photoevaporation achieves high mass-loss rates up to $\sim$10$^{-8}$ {\msunyr} \citep{gorti09a, owen11}. Note that the mass-loss rates can change based on model assumptions such as chemical coolants or the adopted  ultraviolet and X-ray spectra \citep[][]{pascucci22}. Initially, viscous evolution dominates disk evolution, leading to a decrease in the mass accretion rate with time as disk material is depleted onto the star (i.e., less gas in the disk leads to a slower accretion rate onto the star). Once the rate of mass accretion inward through the disk equals the outward photoevaporative mass-loss rate, the inner disk cannot be replenished and is accreted onto the star in $<10^{5}$ yrs \citep{clarke01}.

The onset of rapid disk clearing is reached later for the EUV photoevaporation model since it takes longer for the disk accretion rate to decline to $\sim10^{-10}$ {\msunyr} \citep{hartmann98}. In the case of X-ray--driven photoevaporation, disks should quickly dissipate after reaching a mass accretion rate of $\sim10^{-8}$ {\msunyr}. The average mass accretion rate of young ($\sim1$ Myr old) low-mass pre-main sequence stars (i.e., T~Tauri stars; TTS) with disks is $\sim10^{-8}$ {\msunyr} \citep{hartmann98}, suggesting that X-ray--driven photoevaporation often dominates disk clearing. However, the existence of young stars surrounded by substantial disks with mass accretion rates $<10^{-10}$ {\msunyr} indicates that X-ray--driven photoevaporation may not always dominate \citep{ingleby11a,thanathibodee23}. The observed disk frequency can also be explained by low photoevaporation rates \citep{manzo20}. The main obstacle in distinguishing between photoevaporation models is the lack of observational constraints on the high-energy emission of stars, particularly since our knowledge of the strength of EUV emission from TTS is limited due to absorption of EUV photons by neutral hydrogen along the line of sight.

Infrared (IR) fine structure lines can determine the high-energy radiation fields impinging on the disk surface, which can be used to distinguish between photoevaporation models. Many theoretical works have examined the importance of the neon line luminosities and their connection to high-energy radiation \citep{glassgold07,meijerink08, ercolano08, gorti08, hollenbach09, schisano10, ercolano10}. Due to the high first and second ionization potentials of neon (21.56~eV and 41.0~eV, respectively), the [\ion{Ne}{2}] and [\ion{Ne}{3}] lines lie $>1,000$~K above the ground state, making them ideal to probe high-energy radiation fields.

Both EUV and X-ray radiation have been proposed to be responsible for neon forbidden lines in disks around TTS \citep{glassgold07, hollenbach09}, while far-ultraviolet photons have insufficient energy to ionize neon \citep{glassgold07}. [\ion{Ne}{2}] was detected with the Infrared Spectrograph (IRS) on {\it Spitzer} in more than 50 TTS \citep{espaillat07a, pascucci07, lahuis07, flaccomio09, gudel10, baldovin11, szulagyi12}. However, comparing observations and theories has not led to clear results. High-resolution ground-based observations of the [\ion{Ne}{2}] line are fit equally well by different photo\-evaporation models \citep[e.g.,][]{pascucci11}. 

One way to distinguish between EUV and X-ray creation of neon fine-structure emission is by measuring the [\ion{Ne}{3}]-to-[\ion{Ne}{2}] line flux ratio. A hard EUV spectrum ($L_{\nu}$ $\propto$ $\nu^{-1}$) can lead to [\ion{Ne}{3}]-to-[\ion{Ne}{2}] line flux ratios $>1$ \citep[see Figure 1 in][]{hollenbach09}. Neon ion production by X-rays and/or a soft EUV spectrum can lead to [\ion{Ne}{3}]-to-[\ion{Ne}{2}] line flux ratios of the order $\sim0.1$ \citep{glassgold07, meijerink08}. Of the more than 50 disks with [\ion{Ne}{2}] line detections, only five have [\ion{Ne}{3}] line detections; of these five [\ion{Ne}{3}] line detections, four have [\ion{Ne}{3}]/[\ion{Ne}{2}]$<0.1$ \citep{lahuis07, najita10, espaillat13}. One of the five disks (SZ~Cha) has [\ion{Ne}{3}]/[\ion{Ne}{2}] $>1$, showing that EUV radiation can dominate the creation of neon forbidden lines \citep{espaillat13}. 

SZ~Cha is a TTS with a spectral type of  K0--K2 \citep{rydgren80,manara14}, a stellar accretion rate of $\sim0.2$--$1.6\times10^{-8}$ {\msunyr} \citep{espaillat11, manara14, espaillat19a}, and an X-ray luminosity of $\sim1\times10^{30}$ erg s$^{-1}$ \citep{ingleby11a, espaillat19a}. The IR spectral energy distribution (SED) indicates that SZ~Cha is surrounded by a pre-transitional disk \citep[i.e., an inner disk separated from an outer disk by a large gap $\sim$\,tens of astronomical units wide;][]{kim09, espaillat14}, and SED modeling points to a gap size of about 20~au \citep{espaillat11,ribas16}. SPHERE scattered-light near-infrared images detect disk rings at 40, 70, and 110~au from the star \citep{asensio-torres2021}. Lastly, SZ~Cha displays ``seesaw'' variability in the mid-infrared (MIR) as seen with {\it Spitzer} \citep{espaillat11}. Seesaw variability is characterized by a decrease in emission at shorter IR wavelengths with a corresponding increase at longer IR wavelengths and visa versa \citep{muzerolle09, espaillat11, flaherty12}; this behavior has been attributed to variable shadowing of the outer disk by the inner disk \citep{espaillat11}.

Here we analyze a {\it JWST} spectrum of SZ~Cha as well as a CHIRON optical spectrum taken within 15 hours of the {\it JWST} observations in 2023. We compare the {\it JWST} data to a {\it Spitzer} spectrum of SZ~Cha taken in 2008. In Section~2, we present the new MIR and optical spectra as well as review an archival {\it Spitzer} spectrum and several archival optical spectra of SZ~Cha. In Section~3, we measure the MIR atomic emission lines and model the optical {\halpha} emission line. In Section 4, we compare the 2023 and 2008 data sets and interpret the results in the context of theoretical models. We end with a summary and conclusions in Section~5.

\begin{figure*}   
\epsscale{1.0}
\plotone{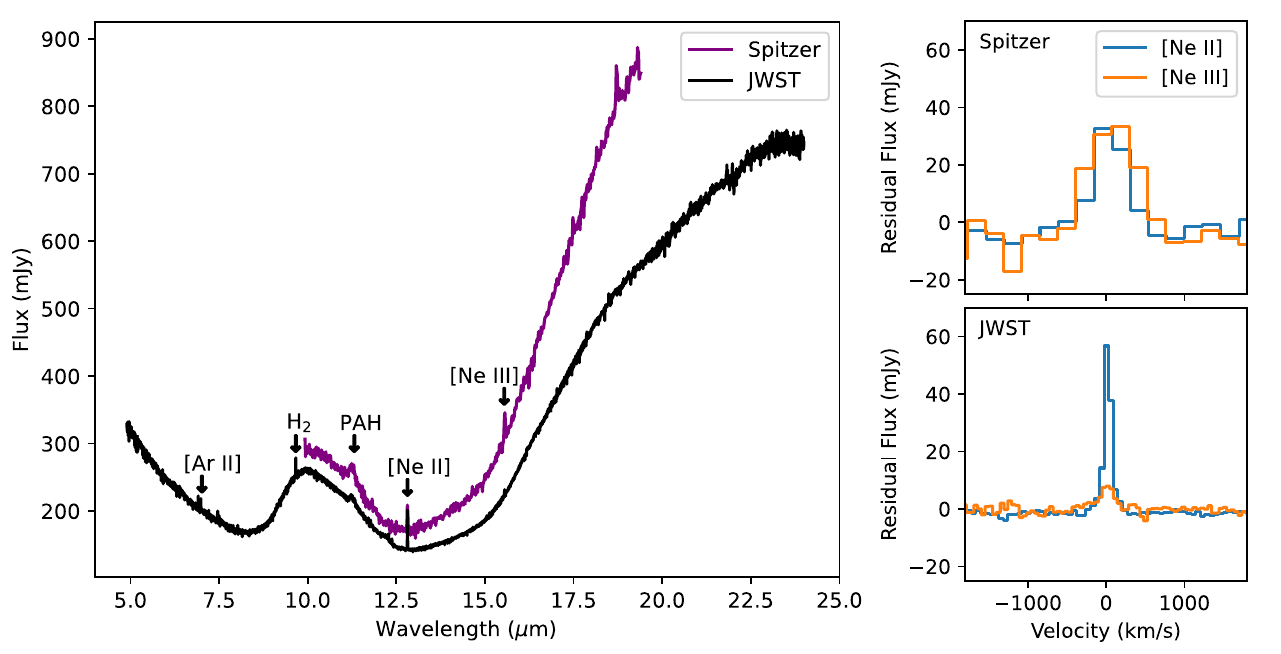} 
\caption{
Left: {\it JWST} MIRI MRS spectrum from 2023 (black) and {\it Spitzer} IRS SH spectrum from 2008 (purple) of SZ~Cha. We label the [\ion{Ar}{2}], H$_2$, [\ion{Ne}{2}], and [\ion{Ne}{3}] emission lines and a polycyclic aromatic hydrocarbon feature around 11 {\micron}. Directly to the left of the [\ion{Ar}{2}] line is an  H$_2$ line. To the right of the [\ion{Ar}{2}] line is a feature where [\ion{He}{2}], [\ion{H}{1}], and [\ion{Ni}{2}] emission lines may overlap.  Right: Comparisons between the {\neii} and {\neiii} lines within the same observation.
}
\end{figure*}    

\section{Observations and Data Reduction} \label{redux}

Here we present a new {\it JWST} Mid-InfraRed Instrument Medium Resolution Spectrometer (MIRI MRS) spectrum of SZ~Cha and a new high-resolution CHIRON optical spectrum taken contemporaneously (within 15~hrs) of the {\it JWST} observations in 2023. In addition, we revisit a previously published {\it Spitzer} IRS spectrum of SZ~Cha from 2008 and archival optical spectra taken at Magellan and ESO between 2007 to 2014.

\begin{figure*}
\epsscale{1.0}
\plotone{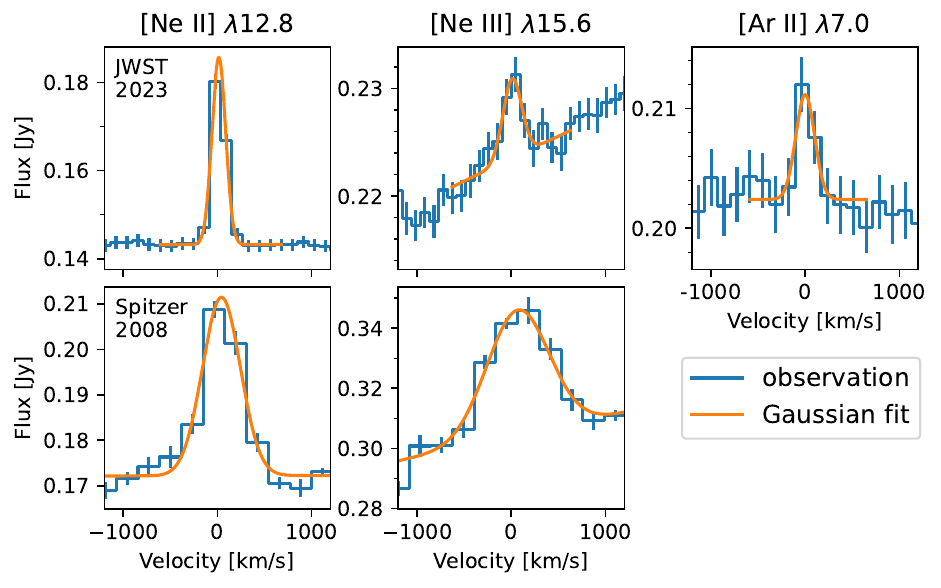}
\caption{
Line fits (orange) to MIR spectra of [\ion{Ne}{2}] 12.81~{\micron}, [\ion{Ne}{3}] 15.55~{\micron}, and [\ion{Ar}{2}] 6.99~{\micron} (blue lines). The top row corresponds to \emph{JWST} and the bottom row to \emph{Spitzer}.
}
\end{figure*} 

\subsection{MIR Spectra}

SZ~Cha was observed with {\it JWST} on 2023 April 21 starting at 10:12 UT using MIRI \citep[][]{reike15, wright23} MRS \citep[][]{wells15} as part of GO program 1676 (PI: C.~Espaillat) in Cycle~1. Dedicated background observations were also obtained. The target and background observations had an exposure time of 336~s and used the default four-point dither. 

We use the {\tt JWST} Science Calibration Pipeline v.1.11.4 \citep{Bushouse2023_jwstpipeline_v1.11.4} to reduce the uncalibrated raw MRS data using the calibration reference file version jwst\_1121.pmap. 
In Stage 1 of the pipeline the raw data or ramps are converted into uncalibrated slope images using the default Detector1 class.
In Stage 2 the dedicated background is subtracted from the slope images, which are subsequently calibrated using the Spec2 class with additional residual fringe correction.
Finally, the composite data cube from all dithered exposures is created from the calibrated slope images in Stage 3. 

The final reduced spectrum is presented in Figure~1 (left) along with a {\it Spitzer} IRS Short-High (SH) spectrum from GO Program 40247, which was published in \citet{espaillat13}. There are significant differences in the continuum level between the \emph{JWST} spectrum and the {\it Spitzer} IRS SH spectrum.  These differences are especially large in channel~4 (17.7--27.9~{\micron}). We do not show data beyond about 24~{\micron}, at which point the spectrum appears to be affected by the reported sensitivity decline in MIRI MRS.  As noted earlier, the MIR continuum of SZ~Cha is known to be variable, but further analysis of channel~4 is necessary to disentangle intrinsic variability from flux calibration issues and we will address this in future work. In Figure~1 (right), we highlight the change in the Neon line ratio between the {\it Spitzer} (top) and {\it JWST} observations (bottom).

\subsection{Optical Spectra}

A high-resolution ($R\sim25,000$) optical spectrum of SZ~Cha was obtained on 2023 April 22 starting at 01:23 UT with the CHIRON bench-mounted, fiber-fed, cross-dispersed echelle spectrograph \citep{schwab10} on a 1.5-m telescope that is part of the Small and Moderate Aperture Research Telescope System \citep{tokovinin13} at Cerro Tololo Inter-American Observatory. The exposure time was 3600~s. The CHIRON observations were taken within $\sim15$ hrs after the \emph{JWST} observations. The data were obtained in fiber mode with $4\times4$ on-chip binning, leading to $R\sim27,800$. The data were reduced following the standard steps using an IDL pipeline.\footnote{http://www.astro.sunysb.edu/fwalter/SMARTS/CHIRON/ch\_reduce.pdf} 

We also searched for archival optical spectra of SZ~Cha. High-resolution ($R\sim35,000$) optical spectra of SZ~Cha were taken with the Magellan Inamori Kyocera Echelle \citep[MIKE;][]{bernstein03} on the Magellan-Clay telescope at Las Campanas Observatory in Chile on the following UT dates: 2007 February 10, February 11; 2008 February 15, February 18; 2009 January 18; 2011 January 5; 2012 February 17. The {\it Spitzer} SH spectrum was taken on 2008 August 16, so we have MIKE spectra taken about six months before (2008 February 15, February 18) the {\it Spitzer} spectrum and another taken about five months after (2009 January 18). The data were reduced using the Image Reduction and Analysis Facility \citep[IRAF;][]{tody93} tasks CDPROC, APFLATTEN, and DOECSLIT. We also retrieved spectra from the ESO Science Archive Facility.\footnote{http://archive.eso.org} The ESO data were taken with X-shooter \citep{vernet11} on 2010 January 19, UVES \citep{dekker00} on 2014 April 17, and FEROS \citep{kaufer99} on 2014 May 31. We present these optical spectra in Section~3.2.

\begin{deluxetable*}{ccccccc}
\tablecaption{MIR Atomic Line Fluxes and Ratios for SZ~Cha \label{Tab:Fluxes}}
\tablehead{
\colhead{Date} \vspace{-0.2cm} & \colhead{Instrument} & \colhead{[\ion{Ar}{2}] 6.99~{\micron}} & \colhead{[\ion{Ne}{2}] 12.81~{\micron}} & \colhead{[\ion{Ne}{3}] 15.55~{\micron}} & \colhead{[\ion{Ne}{3}]-to-[\ion{Ne}{2}]} & \colhead{[\ion{Ne}{2}]-to-[\ion{Ar}{2}]} \\
\colhead{(UT)} & \colhead{} & \colhead{(10$^{-15}$ erg cm$^{-2}$ s$^{-1}$)} & \colhead{(10$^{-15}$ erg cm$^{-2}$ s$^{-1}$)} & \colhead{( 10$^{-15}$erg cm$^{-2}$ s$^{-1}$)} & \colhead{Ratio} & \colhead{Ratio}}
\startdata
2008 Aug 16 &	{\it Spitzer} IRS SH & ... &	$15.5\pm2.9$ & $22.1\pm3.4$ & $1.4\pm0.4$ &	... \\
2023 Apr 21 &	{\it JWST} MIRI MRS	 & $3.2\pm0.6$	& $6.0\pm1.6$ & $1.2\pm0.2$ & $0.20\pm0.06$	& $1.9\pm0.6$
\enddata
\end{deluxetable*}

\begin{figure*}
\epsscale{1.0}
\plotone{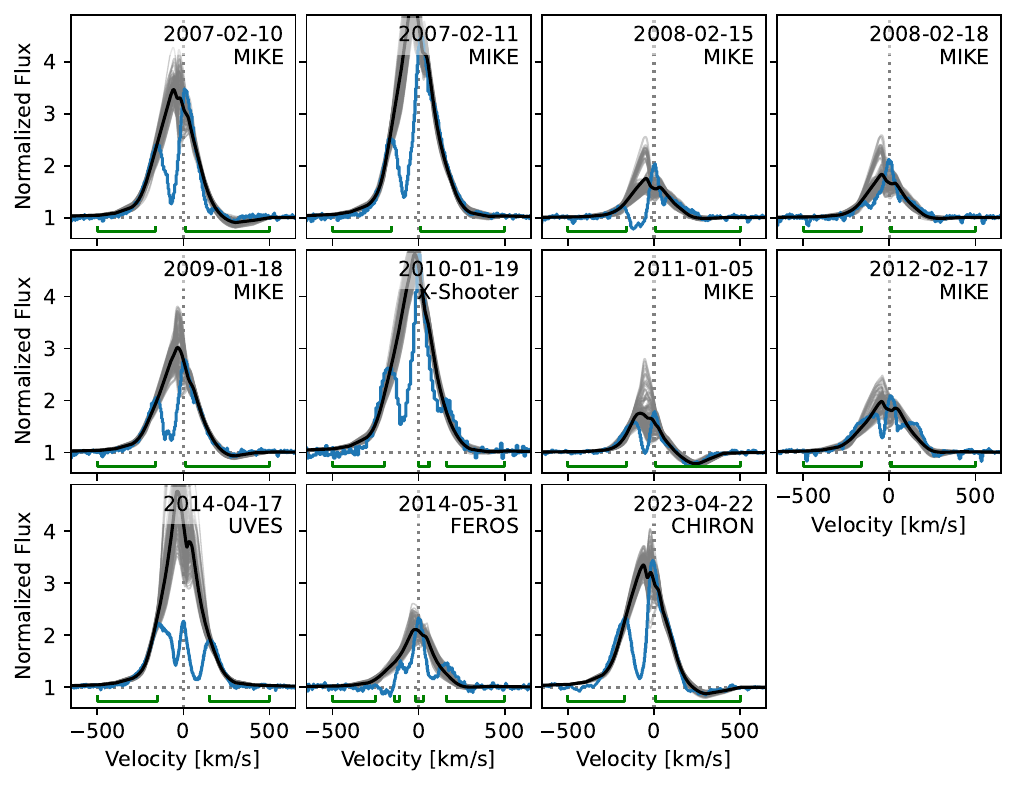} 
\caption{
{\halpha} profiles of SZ~Cha (blue) and the accretion flow modeling fits (black/grey). We show the average accretion flow model fit (black line) from the top 100 best-fitting models (gray lines). The horizontal dotted line denotes the continuum, and the vertical dotted line marks the line center. Horizontal green lines mark regions included in the fit.
}
\end{figure*} 

\section{Analysis and Results} 

In Section~3.1, we identify and measure MIR atomic emission lines. No evidence of any strong molecular emission besides H$_{2}$ in SZ~Cha has emerged. In Section~3.2, we model the {\halpha} emission lines in the CHIRON optical spectra to obtain mass accretion rates.

\subsection{MIR Atomic Lines}

As depicted in Figure~1, we detect the following atomic lines in the \emph{JWST} spectrum: [\ion{Ar}{2}] 6.99~{\micron}, [\ion{Ne}{2}] 12.81~{\micron}, and [\ion{Ne}{3}] 15.55~{\micron}. We note that [\ion{S}{3}] 18.71~{\micron} and [\ion{Ar}{3}] 21.83~{\micron} appear to be present as well, but these are located in channel~4, and due to the issues noted in Section~2.1, we defer analysis of these lines to future work. 

We fit each line and its underlying continuum with a composite model of a Gaussian and a linear function within $\pm700$\,\kms\ from the line center using SciPy's \texttt{curve\_fit}. The fits to the [\ion{Ar}{2}], [\ion{Ne}{2}], and [\ion{Ne}{3}] lines in the \emph{JWST} data are shown in the top panels of Figure~2. Line fluxes, calculated from the width and the amplitude of the Gaussians, and line flux ratios are reported in Table~1. The uncertainty in the fluxes is calculated by propagating the error of the fitted Gaussian parameters returned from \texttt{curve\_fit}. The uncertainties of the fluxes and ratios are calculated following the standard error propagation.

We revisit a \emph{Spitzer} SH spectrum taken on 2008 August 16 and measure the [\ion{Ne}{2}] and [\ion{Ne}{3}] lines (bottom panels of Figure~2, Table~1). Note that the wavelength range of \emph{Spitzer} SH does not cover the [\ion{Ar}{2}] line. We measure neon line fluxes (Table~1) that are consistent with previous work \citep{espaillat13}. 

\subsection{{\halpha} Emission Lines}

In Figure~3 we show {\halpha} emission line profiles of SZ~Cha obtained between 2007 to 2023.  There is significant variability in the {\halpha} emission line strength and in many epochs there is strong blueshifted absorption, likely from a strong, moderate-velocity wind.

Here we model all of the {\halpha} profiles using the magnetospheric accretion flow code from \citet{hartmann94,muzerolle98,muzerolle01}. These models assume alignment of the magnetic, stellar rotation and disk rotation axes. An axisymmetric accretion flow arises from the gas disk that deposits material onto the star. The flow geometry follows that of a dipolar magnetic field and is characterized by an inner radius (\ri) and the width of the flow (\rw) at the disk plane. Additional parameters include the maximum temperature in the flow (\tmax) and the given viewing inclination ($i$).

We created a grid of 133,848 models by varying \mdot~($1-7\times10^{-9}$\,\msunyr), 
\ri~($1.6-6.0$\,R$_{*}$), \rw~($0.2-2.2$\,R$_{*}$), \tmax~($8-11\times10^3$\,K), and $i$~($20\degr-70\degr$). 
We convolved the model profiles with a Gaussian instrumental profile of each instrument's resolution.
We follow the procedure of \citet{thanathibodee23} to determine the best fits and the average model parameters for each observation.  Specifically, we calculate the $\chi^2$ for each combination of the model and observed profile. For each observed profile, we selected the models where the normalized likelihood is $\geq0.5$ and
calculated the weighted mean of \mdot, \ri, \rw, \tmax, and $i$.
We show the best-fitting model to the {\halpha} profiles in Figure~3 and the best fit parameters in Table~2.
Winds are not included in the model; the regions displaying blueshifted absorption and low-velocity redshifted absorption in 2014 were excluded from the fit.

\begin{deluxetable*}{ccccccc}
\tablecaption{Results of the Magnetospheric Accretion Model \label{tab:model_results_acc}}
\tablehead{
\colhead{Obs. Date} &
\colhead{Instrument} &
\colhead{\mdot} &
\colhead{\ri} &
\colhead{\rw} &
\colhead{\tmax} &
\colhead{$i$} \\
\colhead{(UT)} &
\colhead{} &
\colhead{($10^{-9}$\,\msunyr)} &
\colhead{(R$_{\star}$)} &
\colhead{(R$_{\star}$)} &
\colhead{($10^3\,$K)} &
\colhead{(deg)}
}
\startdata
2007-02-10 & MIKE   & 3.8$\pm$1.8 & 2.3$\pm$0.5 & 1.3$\pm$0.5 & 10.6$\pm$0.3 & 65$\pm$ 4 \\ 
2007-02-11 & MIKE   & 4.7$\pm$1.5 & 2.4$\pm$0.5 & 0.7$\pm$0.4 &  9.8$\pm$0.8 & 53$\pm$ 8 \\ 
2008-02-15 & MIKE\tablenotemark{a}   & 3.3$\pm$1.8 & 2.1$\pm$1.3 & 0.5$\pm$0.5 &  9.1$\pm$0.8 & 54$\pm$16 \\ 
2008-02-18 & MIKE\tablenotemark{a}   & 3.2$\pm$1.8 & 2.1$\pm$1.2 & 0.5$\pm$0.5 &  9.3$\pm$0.8 & 51$\pm$16 \\ 
2009-01-18 & MIKE\tablenotemark{b}   & 4.4$\pm$1.8 & 2.0$\pm$0.4 & 0.8$\pm$0.4 & 10.1$\pm$0.7 & 58$\pm$10 \\ 
2010-01-19 & X-Shooter     & 4.8$\pm$1.6 & 2.4$\pm$0.5 & 0.9$\pm$0.6 & 10.1$\pm$0.7 & 57$\pm$11 \\ 
2011-01-05 & MIKE   & 3.5$\pm$1.8 & 3.1$\pm$1.5 & 1.2$\pm$0.5 &  9.6$\pm$0.8 & 39$\pm$17 \\ 
2012-02-17 & MIKE   & 4.3$\pm$1.8 & 1.8$\pm$0.6 & 0.4$\pm$0.3 &  9.3$\pm$0.8 & 58$\pm$11 \\ 
2014-04-17 & UVES   & 4.0$\pm$1.4 & 2.2$\pm$0.3 & 0.6$\pm$0.4 &  9.7$\pm$0.8 & 44$\pm$10 \\ 
2014-05-31 & FEROS  & 2.7$\pm$1.4 & 1.6$\pm$0.1 & 0.2$\pm$0.0 & 10.0$\pm$0.6 & 61$\pm$ 9 \\ 
2023-04-22 & CHIRON\tablenotemark{c} & 5.2$\pm$1.3 & 2.1$\pm$0.2 & 1.6$\pm$0.4 & 10.6$\pm$0.4 & 65$\pm$ 3 \\ 
\enddata
\tablenotetext{a}{Observation taken about 6 months before the {\it Spitzer} observation.}
\tablenotetext{b}{Observation taken about 5 months after the {\it Spitzer} observation.}
\tablenotetext{c}{Observation taken about 15 hrs after the {\it JWST} observation.}
\end{deluxetable*}

\section{Discussion} \label{sec:discussion}

We do not detect any significant molecular emission besides H$_{2}$ from the pre-transitional disk of SZ~Cha. This is consistent with previous MIR molecular line observations of transitional disks \citep[i.e., a disk with a central cavity devoid of dust and $\sim$\,tens of astronomical units wide;][]{espaillat14} that find no significant molecular emission \citep{najita10, banzatti20}; this is in contrast to observations of disks without large cavities (i.e., ``full'' disks) that are rich in molecular emission \citep{carr08, pontoppidan10, banzatti23}. 

Pre-transitional and transitional disks are ideal candidates to search for [\ion{Ne}{3}] line emission. Only five protoplanetary disks have [\ion{Ne}{3}] line detections to date \citep{lahuis07, najita10, espaillat13}. Of these, four were from pre-transitional and transitional disks. As noted above, little molecular emission is found in pre-transitional and transitional disks, which helps with the detection of the [Ne III] line since there are H$_2$O lines in the [Ne III] wavelength region. In addition, there is less MIR continuum emission in pre-transitional and transitional disks, which facilitates line detection. Also, pre-transitional and transitional disks are not known to have strong outflows or jets as seen in full disks that can contribute to neon emission \citep{gudel10}. High-resolution ground-based MIR spectroscopy shows that the [Ne II] line emission comes from a region close ($\sim$\,tens of astronomical units) to the star for pre-transitional and transitional disks \citep{sacco12}.

Aside from the {\it Spitzer} SH observation of SZ~Cha, no evidence has been found to support the possibility that EUV radiation dominates the creation of MIR atomic emission lines. \citet{szulagyi12} have studied a sample of $\sim60$ pre-transitional and transitional disks with {\it Spitzer} and find [\ion{Ne}{3}]-to-[\ion{Ne}{2}] line flux ratios $<0.1$, which point to a soft EUV or X-ray spectrum. Indirect measurements of the EUV luminosity in TTS using radio emission suggest that the EUV emission is not sufficient to reproduce the observed [\ion{Ne}{2}] 12.81~{\micron} luminosities \citep{pascucci14, pascucci20}. 

To explain the large difference between the [\ion{Ne}{3}]-to-[\ion{Ne}{2}] line flux ratio measured by {\it Spitzer} and {\it JWST}, we consider the possibility that the {\it Spitzer} measurement of the [\ion{Ne}{3}] line flux of SZ~Cha is contaminated by unresolved line emission. As a test, we convolved our \emph{JWST} spectrum down to the resolution of {\it Spitzer} IRS SH and see that the [\ion{Ne}{3}]-to-[\ion{Ne}{2}] line flux ratio does not change. Therefore, it is unlikely that line contamination played a significant role in the [\ion{Ne}{3}] line flux ratio measurement of SZ~Cha obtained with {\it Spitzer}. We then consider if at the time of the {\it Spitzer} observations strong emission existed from the 13$_{6~8}$ to 12$_{3~9}$ transition of H$_2$O, which is located at $\sim15.57$~{\micron} and may be weakly present in the {\it JWST} spectrum (see top middle panel of Figure~2). However, this is unlikely given that no water emission was seen in the {\it Spitzer} SH spectrum at any other wavelengths.

Next we consider that the significant change in the [\ion{Ne}{3}]-to-[\ion{Ne}{2}] line flux ratio of SZ~Cha between 2008 and 2023 is due to a jet that was present at the time of the {\it Spitzer} observations and that is no longer active at the time of the {\it JWST} observations. \citet{pascucci14} did not find evidence of a jet when studying centimeter emission of SZ~Cha in 2012 October. The variability of jet emission is not currently well understood \citep{espaillat19b, espaillat22, curone23}, and while we cannot rule out this scenario, we consider it to be unlikely.

To explain the variability in the [\ion{Ne}{3}]-to-[\ion{Ne}{2}] line flux ratio of SZ~Cha between 2008 and 2023, we propose a variable wind. A signature of a wind is blueshifted absorption in the {\halpha} profile, and in Figure~3 it is clear that SZ~Cha's wind is variable on short timescales (i.e., days). The presence of strong blueshifted absorption is evident in the {\halpha} profile of SZ~Cha that was obtained about 15 hrs after the {\it JWST} observations (bottom row, right panel, Figure~3). To the best of our knowledge, no {\halpha} observations of SZ~Cha have been taken contemporaneously with {\it Spitzer}. There are MIKE spectra taken about six months before (top row, two right panels in Figure~3) and about five months after (second row, leftmost panel in Figure~3) the {\it Spitzer} observations. We propose that at the time of the {\it Spitzer} observations, the wind was suppressed, and that during the {\it JWST} observations, a strong wind existed. A strong wind would absorb the EUV from the star and prevent EUV emission from reaching the disk, but X-rays would permeate the wind and reach the disk and generate the neon emission.

We can consider the [\ion{Ne}{2}]-to-[\ion{Ar}{2}] line flux ratio to further investigate if a strong wind was present at the time of the {\it JWST} observations. \citet{hollenbach09} find that [\ion{Ne}{2}]/[\ion{Ar}{2}]$\sim$1 in the case that EUV or soft X-ray photons generate these emission lines and that this ratio is about 2.5 in the case of hard X-rays.    
The measured [\ion{Ne}{2}]-to-[\ion{Ar}{2}] line flux ratio of 1.9$\pm$0.6 is roughly compatible with either scenario within the uncertainties. More {\it JWST} data are necessary to achieve a better signal-to-noise ratio.  Additional contemporaneous multiwavelength observations of SZ~Cha can test how the [\ion{Ne}{3}]-to-[\ion{Ne}{2}] line flux ratio varies with the wind.

\section{Summary and Conclusions} 

We present a 2023 {\it JWST} MIRI MRS spectrum of the pre-transitional disk of SZ~Cha. We find no significant molecular emission, in line with previous {\it Spitzer} IRS observations of this object specifically and disks with large gaps in general. We detect atomic argon and neon emission, particularly the [\ion{Ar}{2}] 6.99~{\micron}, [\ion{Ne}{2}] 12.81~{\micron}, and [\ion{Ne}{3}] 15.55~{\micron} fine-structure emission lines. The {\it JWST} [\ion{Ne}{3}]-to-[\ion{Ne}{2}] line flux ratio point to X-ray--dominated photoevaporation of the disk.

We compare the neon line flux ratios from {\it JWST} in 2023 to those measured with {\it Spitzer} in 2008 and find significant variability. With {\it Spitzer}, the [\ion{Ne}{3}]-to-[\ion{Ne}{2}] line flux ratio pointed to EUV-dominated photoevaporation of the disk. The change in the [\ion{Ne}{3}]-to-[\ion{Ne}{2}] line flux ratio between the {\it Spitzer} and {\it JWST} observations indicates a switch between EUV and X-ray disk photoevaporation. To explore the reason behind this switch, we analyze a ground-based optical spectrum of the {\halpha} emission line taken within 15 hrs of the {\it JWST} observations. We find strong blueshifted absorption, evidence of a wind. We propose that a strong wind was present at the time of the {\it JWST} observations and that the EUV radiation was prevented from reaching the disk while X-rays were able to permeate the wind and irradiate the disk. By extension, we propose that a weak wind was present at the time of the {\it Spitzer} observations and therefore EUV photons reached the disk and dominated the disk photoevaporation.

This study shows that disk photoevaporation can be dominated by either EUV or X-ray radiation and that this is impacted by the strength of winds that play a role in dictating how much EUV reaches the disk. More observations of MIR atomic lines in a larger sample of protoplanetary disks are necessary to better understand how MIR atomic emission can be used to study disk photoevaporation. In addition, more expansive multiepoch multiwavelength contemporaneous data sets with {\it JWST} are critical to probe the star-disk connection.

 \acknowledgments{
We thank Todd Henry, Wei-Chun Jao, and the CHIRON team for their prompt scheduling of the CHIRON observations. We acknowledge support from {\it JWST} grant GO-01676. M.K.M. acknowledges support from the Dutch Research Council (NWO; grant VI.Veni.192.241).

The JWST data presented in this paper were obtained from the Mikulski Archive for Space Telescopes (MAST) at the Space Telescope Science Institute. The specific observations analyzed can be accessed via \dataset[10.17909/092y-8624]{https://doi.org/10.17909/092y-8624}.
}


\begin{thebibliography}{}
\expandafter\ifx\csname natexlab\endcsname\relax\def\natexlab#1{#1}\fi
\providecommand{\url}[1]{\href{#1}{#1}}
\providecommand{\dodoi}[1]{doi:~\href{http://doi.org/#1}{\nolinkurl{#1}}}
\providecommand{\doeprint}[1]{\href{http://ascl.net/#1}{\nolinkurl{http://ascl.net/#1}}}
\providecommand{\doarXiv}[1]{\href{https://arxiv.org/abs/#1}{\nolinkurl{https://arxiv.org/abs/#1}}}

\bibitem[{{Alexander} {et~al.}(2006){Alexander}, {Clarke}, \&
  {Pringle}}]{alexander06}
{Alexander}, R.~D., {Clarke}, C.~J., \& {Pringle}, J.~E. 2006, \mnras, 369,
  229, \dodoi{10.1111/j.1365-2966.2006.10294.x}

\bibitem[{{Asensio-Torres} {et~al.}(2021){Asensio-Torres}, {Henning},
  {Cantalloube}, {Pinilla}, {Mesa}, {Garufi}, {Jorquera}, {Gratton}, {Chauvin},
  {Szul{\'a}gyi}, {van Boekel}, {Dong}, {Marleau}, {Benisty}, {Villenave},
  {Bergez-Casalou}, {Desgrange}, {Janson}, {Keppler}, {Langlois}, {M{\'e}nard},
  {Rickman}, {Stolker}, {Feldt}, {Fusco}, {Gluck}, {Pavlov}, \&
  {Ramos}}]{asensio-torres2021}
{Asensio-Torres}, R., {Henning}, T., {Cantalloube}, F., {et~al.} 2021, \aap,
  652, A101, \dodoi{10.1051/0004-6361/202140325}

\bibitem[{{Baldovin-Saavedra} {et~al.}(2011){Baldovin-Saavedra}, {Audard},
  {G{\"u}del}, {Rebull}, {Padgett}, {Skinner}, {Carmona}, {Glauser}, \&
  {Fajardo-Acosta}}]{baldovin11}
{Baldovin-Saavedra}, C., {Audard}, M., {G{\"u}del}, M., {et~al.} 2011, \aap,
  528, A22, \dodoi{10.1051/0004-6361/201015622}

\bibitem[{{Banzatti} {et~al.}(2020){Banzatti}, {Pascucci}, {Bosman}, {Pinilla},
  {Salyk}, {Herczeg}, {Pontoppidan}, {Vazquez}, {Watkins}, {Krijt}, {Hendler},
  \& {Long}}]{banzatti20}
{Banzatti}, A., {Pascucci}, I., {Bosman}, A.~D., {et~al.} 2020, \apj, 903, 124,
  \dodoi{10.3847/1538-4357/abbc1a}

\bibitem[{{Banzatti} {et~al.}(2023){Banzatti}, {Pontoppidan}, {Carr},
  {Jellison}, {Pascucci}, {Najita}, {Munoz-Romero}, {Oberg}, {Kalyaan},
  {Pinilla}, {Krijt}, {Long}, {Lambrechts}, {Rosotti}, {Herczeg}, {Salyk},
  {Zhang}, {Ballering}, {Meyer}, {Bruderer}, \& {the JDISCS
  collaboration}}]{banzatti23}
{Banzatti}, A., {Pontoppidan}, K.~M., {Carr}, J., {et~al.} 2023, arXiv
  e-prints, arXiv:2307.03846, \dodoi{10.48550/arXiv.2307.03846}

\bibitem[{{Bernstein} {et~al.}(2003){Bernstein}, {Shectman}, {Gunnels},
  {Mochnacki}, \& {Athey}}]{bernstein03}
{Bernstein}, R., {Shectman}, S.~A., {Gunnels}, S.~M., {Mochnacki}, S., \&
  {Athey}, A.~E. 2003, in Society of Photo-Optical Instrumentation Engineers
  (SPIE) Conference Series, Vol. 4841, Society of Photo-Optical Instrumentation
  Engineers (SPIE) Conference Series, ed. {M.~Iye \& A.~F.~M.~Moorwood},
  1694--1704, \dodoi{10.1117/12.461502}

\bibitem[{Bushouse {et~al.}(2023)Bushouse, Eisenhamer, Dencheva, Davies,
  Greenfield, Morrison, Hodge, Simon, Grumm, Droettboom, Slavich, Sosey, Pauly,
  Miller, Jedrzejewski, Hack, Davis, Crawford, Law, Gordon, Regan, Cara,
  MacDonald, Bradley, Shanahan, Jamieson, Teodoro, \&
  Williams}]{Bushouse2023_jwstpipeline_v1.11.4}
Bushouse, H., Eisenhamer, J., Dencheva, N., {et~al.} 2023, JWST Calibration
  Pipeline, 1.11.4,  Zenodo, \dodoi{10.5281/zenodo.8247246}

\bibitem[{{Carr} \& {Najita}(2008)}]{carr08}
{Carr}, J.~S., \& {Najita}, J.~R. 2008, Science, 319, 1504,
  \dodoi{10.1126/science.1153807}

\bibitem[{{Clarke} {et~al.}(2001){Clarke}, {Gendrin}, \&
  {Sotomayor}}]{clarke01}
{Clarke}, C.~J., {Gendrin}, A., \& {Sotomayor}, M. 2001, \mnras, 328, 485,
  \dodoi{10.1046/j.1365-8711.2001.04891.x}

\bibitem[{{Curone} {et~al.}(2023){Curone}, {Testi}, {Macias}, {Tazzari},
  {Facchini}, {Williams}, {Clarke}, {Natta}, {Rosotti}, {Toci}, \&
  {Lodato}}]{curone23}
{Curone}, P., {Testi}, L., {Macias}, E., {et~al.} 2023, arXiv e-prints,
  arXiv:2307.10798, \dodoi{10.48550/arXiv.2307.10798}

\bibitem[{{Dekker} {et~al.}(2000){Dekker}, {D'Odorico}, {Kaufer}, {Delabre}, \&
  {Kotzlowski}}]{dekker00}
{Dekker}, H., {D'Odorico}, S., {Kaufer}, A., {Delabre}, B., \& {Kotzlowski}, H.
  2000, in Society of Photo-Optical Instrumentation Engineers (SPIE) Conference
  Series, Vol. 4008, Optical and IR Telescope Instrumentation and Detectors,
  ed. M.~{Iye} \& A.~F. {Moorwood}, 534--545, \dodoi{10.1117/12.395512}

\bibitem[{{Ercolano} {et~al.}(2008){Ercolano}, {Drake}, {Raymond}, \&
  {Clarke}}]{ercolano08}
{Ercolano}, B., {Drake}, J.~J., {Raymond}, J.~C., \& {Clarke}, C.~C. 2008,
  \apj, 688, 398, \dodoi{10.1086/590490}

\bibitem[{{Ercolano} \& {Owen}(2010)}]{ercolano10}
{Ercolano}, B., \& {Owen}, J.~E. 2010, \mnras, 406, 1553,
  \dodoi{10.1111/j.1365-2966.2010.16798.x}

\bibitem[{{Espaillat} {et~al.}(2011){Espaillat}, {Furlan}, {D'Alessio},
  {Sargent}, {Nagel}, {Calvet}, {Watson}, \& {Muzerolle}}]{espaillat11}
{Espaillat}, C., {Furlan}, E., {D'Alessio}, P., {et~al.} 2011, \apj, 728, 49,
  \dodoi{10.1088/0004-637X/728/1/49}

\bibitem[{{Espaillat} {et~al.}(2007){Espaillat}, {Calvet}, {D'Alessio},
  {Bergin}, {Hartmann}, {Watson}, {Furlan}, {Najita}, {Forrest}, {McClure},
  {Sargent}, {Bohac}, \& {Harrold}}]{espaillat07a}
{Espaillat}, C., {Calvet}, N., {D'Alessio}, P., {et~al.} 2007, \apjl, 664,
  L111, \dodoi{10.1086/520879}

\bibitem[{{Espaillat} {et~al.}(2013){Espaillat}, {Ingleby}, {Furlan},
  {McClure}, {Spatzier}, {Nieusma}, {Calvet}, {Bergin}, {Hartmann}, {Miller},
  \& {Muzerolle}}]{espaillat13}
{Espaillat}, C., {Ingleby}, L., {Furlan}, E., {et~al.} 2013, \apj, 762, 62,
  \dodoi{10.1088/0004-637X/762/1/62}

\bibitem[{{Espaillat} {et~al.}(2014){Espaillat}, {Muzerolle}, {Najita},
  {Andrews}, {Zhu}, {Calvet}, {Kraus}, {Hashimoto}, {Kraus}, \&
  {D'Alessio}}]{espaillat14}
{Espaillat}, C., {Muzerolle}, J., {Najita}, J., {et~al.} 2014, in Protostars
  and Planets VI, ed. H.~{Beuther}, R.~S. {Klessen}, C.~P. {Dullemond}, \&
  T.~{Henning}, 497, \dodoi{10.2458/azu_uapress_9780816531240-ch022}

\bibitem[{{Espaillat} {et~al.}(2019{\natexlab{a}}){Espaillat}, {Mac{\'\i}as},
  {Hern{\'a}ndez}, \& {Robinson}}]{espaillat19b}
{Espaillat}, C.~C., {Mac{\'\i}as}, E., {Hern{\'a}ndez}, J., \& {Robinson}, C.
  2019{\natexlab{a}}, \apjl, 877, L34, \dodoi{10.3847/2041-8213/ab2193}

\bibitem[{{Espaillat} {et~al.}(2022){Espaillat}, {Mac{\'\i}as}, {Wendeborn},
  {Franco-Hern{\'a}ndez}, {Calvet}, {Rilinger}, {Cleeves}, \&
  {D'Alessio}}]{espaillat22}
{Espaillat}, C.~C., {Mac{\'\i}as}, E., {Wendeborn}, J., {et~al.} 2022, \apj,
  924, 104, \dodoi{10.3847/1538-4357/ac365a}

\bibitem[{{Espaillat} {et~al.}(2019{\natexlab{b}}){Espaillat}, {Robinson},
  {Grant}, \& {Reynolds}}]{espaillat19a}
{Espaillat}, C.~C., {Robinson}, C., {Grant}, S., \& {Reynolds}, M.
  2019{\natexlab{b}}, \apj, 876, 121, \dodoi{10.3847/1538-4357/ab16e6}

\bibitem[{{Flaccomio} {et~al.}(2009){Flaccomio}, {Stelzer}, {Sciortino},
  {Micela}, {Pillitteri}, \& {Testi}}]{flaccomio09}
{Flaccomio}, E., {Stelzer}, B., {Sciortino}, S., {et~al.} 2009, \aap, 505, 695,
  \dodoi{10.1051/0004-6361/200810972}

\bibitem[{{Flaherty} {et~al.}(2012){Flaherty}, {Muzerolle}, {Rieke},
  {Gutermuth}, {Balog}, {Herbst}, {Megeath}, \& {Kun}}]{flaherty12}
{Flaherty}, K.~M., {Muzerolle}, J., {Rieke}, G., {et~al.} 2012, \apj, 748, 71,
  \dodoi{10.1088/0004-637X/748/1/71}

\bibitem[{{Font} {et~al.}(2004){Font}, {McCarthy}, {Johnstone}, \&
  {Ballantyne}}]{font04}
{Font}, A.~S., {McCarthy}, I.~G., {Johnstone}, D., \& {Ballantyne}, D.~R. 2004,
  \apj, 607, 890, \dodoi{10.1086/383518}

\bibitem[{{Glassgold} {et~al.}(2007){Glassgold}, {Najita}, \&
  {Igea}}]{glassgold07}
{Glassgold}, A.~E., {Najita}, J.~R., \& {Igea}, J. 2007, \apj, 656, 515,
  \dodoi{10.1086/510013}

\bibitem[{{Gorti} \& {Hollenbach}(2008)}]{gorti08}
{Gorti}, U., \& {Hollenbach}, D. 2008, \apj, 683, 287, \dodoi{10.1086/589616}

\bibitem[{{Gorti} \& {Hollenbach}(2009)}]{gorti09a}
---. 2009, \apj, 690, 1539, \dodoi{10.1088/0004-637X/690/2/1539}

\bibitem[{{G{\"u}del} {et~al.}(2010){G{\"u}del}, {Lahuis}, {Briggs}, {Carr},
  {Glassgold}, {Henning}, {Najita}, {van Boekel}, \& {van Dishoeck}}]{gudel10}
{G{\"u}del}, M., {Lahuis}, F., {Briggs}, K.~R., {et~al.} 2010, \aap, 519, A113,
  \dodoi{10.1051/0004-6361/200913971}

\bibitem[{{Hartmann} {et~al.}(1998){Hartmann}, {Calvet}, {Gullbring}, \&
  {D'Alessio}}]{hartmann98}
{Hartmann}, L., {Calvet}, N., {Gullbring}, E., \& {D'Alessio}, P. 1998, \apj,
  495, 385, \dodoi{10.1086/305277}

\bibitem[{{Hartmann} {et~al.}(1994){Hartmann}, {Hewett}, \&
  {Calvet}}]{hartmann94}
{Hartmann}, L., {Hewett}, R., \& {Calvet}, N. 1994, \apj, 426, 669,
  \dodoi{10.1086/174104}

\bibitem[{{Hollenbach} \& {Gorti}(2009)}]{hollenbach09}
{Hollenbach}, D., \& {Gorti}, U. 2009, \apj, 703, 1203,
  \dodoi{10.1088/0004-637X/703/2/1203}

\bibitem[{{Ingleby} {et~al.}(2011){Ingleby}, {Calvet}, {Hern{\'a}ndez},
  {Brice{\~n}o}, {Espaillat}, {Miller}, {Bergin}, \& {Hartmann}}]{ingleby11a}
{Ingleby}, L., {Calvet}, N., {Hern{\'a}ndez}, J., {et~al.} 2011, \aj, 141, 127,
  \dodoi{10.1088/0004-6256/141/4/127}

\bibitem[{{Kaufer} {et~al.}(1999){Kaufer}, {Stahl}, {Tubbesing},
  {N{\o}rregaard}, {Avila}, {Francois}, {Pasquini}, \& {Pizzella}}]{kaufer99}
{Kaufer}, A., {Stahl}, O., {Tubbesing}, S., {et~al.} 1999, The Messenger, 95, 8

\bibitem[{{Kim} {et~al.}(2009){Kim}, {Watson}, {Manoj}, {Furlan}, {Najita},
  {Forrest}, {Sargent}, {Espaillat}, {Calvet}, {Luhman}, {McClure}, {Green}, \&
  {Harrold}}]{kim09}
{Kim}, K.~H., {Watson}, D.~M., {Manoj}, P., {et~al.} 2009, \apj, 700, 1017,
  \dodoi{10.1088/0004-637X/700/2/1017}

\bibitem[{{Lahuis} {et~al.}(2007){Lahuis}, {van Dishoeck}, {Blake}, {Evans},
  {Kessler-Silacci}, \& {Pontoppidan}}]{lahuis07}
{Lahuis}, F., {van Dishoeck}, E.~F., {Blake}, G.~A., {et~al.} 2007, \apj, 665,
  492, \dodoi{10.1086/518931}

\bibitem[{{Manara} {et~al.}(2014){Manara}, {Testi}, {Natta}, {Rosotti},
  {Benisty}, {Ercolano}, \& {Ricci}}]{manara14}
{Manara}, C.~F., {Testi}, L., {Natta}, A., {et~al.} 2014, \aap, 568, A18,
  \dodoi{10.1051/0004-6361/201323318}

\bibitem[{{Manzo-Mart{\'\i}nez} {et~al.}(2020){Manzo-Mart{\'\i}nez}, {Calvet},
  {Hern{\'a}ndez}, {Lizano}, {Hern{\'a}ndez}, {Miller}, {Mauc{\'o}},
  {Brice{\~n}o}, \& {D'Alessio}}]{manzo20}
{Manzo-Mart{\'\i}nez}, E., {Calvet}, N., {Hern{\'a}ndez}, J., {et~al.} 2020,
  \apj, 893, 56, \dodoi{10.3847/1538-4357/ab7ead}

\bibitem[{{Meijerink} {et~al.}(2008){Meijerink}, {Glassgold}, \&
  {Najita}}]{meijerink08}
{Meijerink}, R., {Glassgold}, A.~E., \& {Najita}, J.~R. 2008, \apj, 676, 518,
  \dodoi{10.1086/527411}

\bibitem[{{Muzerolle} {et~al.}(1998){Muzerolle}, {Calvet}, \&
  {Hartmann}}]{muzerolle98}
{Muzerolle}, J., {Calvet}, N., \& {Hartmann}, L. 1998, \apj, 492, 743,
  \dodoi{10.1086/305069}

\bibitem[{{Muzerolle} {et~al.}(2001){Muzerolle}, {Calvet}, \&
  {Hartmann}}]{muzerolle01}
---. 2001, \apj, 550, 944, \dodoi{10.1086/319779}

\bibitem[{{Muzerolle} {et~al.}(2009){Muzerolle}, {Flaherty}, {Balog}, {Furlan},
  {Smith}, {Allen}, {Calvet}, {D'Alessio}, {Megeath}, {Muench}, {Rieke}, \&
  {Sherry}}]{muzerolle09}
{Muzerolle}, J., {Flaherty}, K., {Balog}, Z., {et~al.} 2009, \apjl, 704, L15,
  \dodoi{10.1088/0004-637X/704/1/L15}

\bibitem[{{Najita} {et~al.}(2010){Najita}, {Carr}, {Strom}, {Watson},
  {Pascucci}, {Hollenbach}, {Gorti}, \& {Keller}}]{najita10}
{Najita}, J.~R., {Carr}, J.~S., {Strom}, S.~E., {et~al.} 2010, \apj, 712, 274,
  \dodoi{10.1088/0004-637X/712/1/274}

\bibitem[{{Owen} {et~al.}(2011){Owen}, {Ercolano}, \& {Clarke}}]{owen11}
{Owen}, J.~E., {Ercolano}, B., \& {Clarke}, C.~J. 2011, \mnras, 412, 13,
  \dodoi{10.1111/j.1365-2966.2010.17818.x}

\bibitem[{{Pascucci} {et~al.}(2022){Pascucci}, {Cabrit}, {Edwards}, {Gorti},
  {Gressel}, \& {Suzuki}}]{pascucci22}
{Pascucci}, I., {Cabrit}, S., {Edwards}, S., {et~al.} 2022, arXiv e-prints,
  arXiv:2203.10068, \dodoi{10.48550/arXiv.2203.10068}

\bibitem[{{Pascucci} {et~al.}(2014){Pascucci}, {Ricci}, {Gorti}, {Hollenbach},
  {Hendler}, {Brooks}, \& {Contreras}}]{pascucci14}
{Pascucci}, I., {Ricci}, L., {Gorti}, U., {et~al.} 2014, \apj, 795, 1,
  \dodoi{10.1088/0004-637X/795/1/1}

\bibitem[{{Pascucci} {et~al.}(2007){Pascucci}, {Hollenbach}, {Najita},
  {Muzerolle}, {Gorti}, {Herczeg}, {Hillenbrand}, {Kim}, {Carpenter}, {Meyer},
  {Mamajek}, \& {Bouwman}}]{pascucci07}
{Pascucci}, I., {Hollenbach}, D., {Najita}, J., {et~al.} 2007, \apj, 663, 383,
  \dodoi{10.1086/518535}

\bibitem[{{Pascucci} {et~al.}(2011){Pascucci}, {Sterzik}, {Alexander},
  {Alencar}, {Gorti}, {Hollenbach}, {Owen}, {Ercolano}, \&
  {Edwards}}]{pascucci11}
{Pascucci}, I., {Sterzik}, M., {Alexander}, R.~D., {et~al.} 2011, \apj, 736,
  13, \dodoi{10.1088/0004-637X/736/1/13}

\bibitem[{{Pascucci} {et~al.}(2020){Pascucci}, {Banzatti}, {Gorti}, {Fang},
  {Pontoppidan}, {Alexander}, {Ballabio}, {Edwards}, {Salyk}, {Sacco},
  {Flaccomio}, {Blake}, {Carmona}, {Hall}, {Kamp}, {K{\"a}ufl}, {Meeus},
  {Meyer}, {Pauly}, {Steendam}, \& {Sterzik}}]{pascucci20}
{Pascucci}, I., {Banzatti}, A., {Gorti}, U., {et~al.} 2020, \apj, 903, 78,
  \dodoi{10.3847/1538-4357/abba3c}

\bibitem[{{Pontoppidan} {et~al.}(2010){Pontoppidan}, {Salyk}, {Blake},
  {Meijerink}, {Carr}, \& {Najita}}]{pontoppidan10}
{Pontoppidan}, K.~M., {Salyk}, C., {Blake}, G.~A., {et~al.} 2010, \apj, 720,
  887, \dodoi{10.1088/0004-637X/720/1/887}

\bibitem[{{Ribas} {et~al.}(2016){Ribas}, {Bouy}, {Mer{\'\i}n}, {Duch{\^e}ne},
  {Rebollido}, {Espaillat}, \& {Pinte}}]{ribas16}
{Ribas}, {\'A}., {Bouy}, H., {Mer{\'\i}n}, B., {et~al.} 2016, \mnras, 458,
  1029, \dodoi{10.1093/mnras/stw333}

\bibitem[{{Rieke} {et~al.}(2015){Rieke}, {Wright}, {B{\"o}ker}, {Bouwman},
  {Colina}, {Glasse}, {Gordon}, {Greene}, {G{\"u}del}, {Henning}, {Justtanont},
  {Lagage}, {Meixner}, {N{\o}rgaard-Nielsen}, {Ray}, {Ressler}, {van Dishoeck},
  \& {Waelkens}}]{reike15}
{Rieke}, G.~H., {Wright}, G.~S., {B{\"o}ker}, T., {et~al.} 2015, \pasp, 127,
  584, \dodoi{10.1086/682252}

\bibitem[{{Rydgren}(1980)}]{rydgren80}
{Rydgren}, A.~E. 1980, \aj, 85, 444, \dodoi{10.1086/112694}

\bibitem[{{Sacco} {et~al.}(2012){Sacco}, {Flaccomio}, {Pascucci}, {Lahuis},
  {Ercolano}, {Kastner}, {Micela}, {Stelzer}, \& {Sterzik}}]{sacco12}
{Sacco}, G.~G., {Flaccomio}, E., {Pascucci}, I., {et~al.} 2012, \apj, 747, 142,
  \dodoi{10.1088/0004-637X/747/2/142}

\bibitem[{{Schisano} {et~al.}(2010){Schisano}, {Ercolano}, \&
  {G{\"u}del}}]{schisano10}
{Schisano}, E., {Ercolano}, B., \& {G{\"u}del}, M. 2010, \mnras, 401, 1636,
  \dodoi{10.1111/j.1365-2966.2009.15799.x}

\bibitem[{{Schwab} {et~al.}(2010){Schwab}, {Spronck}, {Tokovinin}, \&
  {Fischer}}]{schwab10}
{Schwab}, C., {Spronck}, J. F.~P., {Tokovinin}, A., \& {Fischer}, D.~A. 2010,
  in Society of Photo-Optical Instrumentation Engineers (SPIE) Conference
  Series, Vol. 7735, \procspie, 77354G, \dodoi{10.1117/12.856709}

\bibitem[{{Szul{\'a}gyi} {et~al.}(2012){Szul{\'a}gyi}, {Pascucci},
  {{\'A}brah{\'a}m}, {Apai}, {Bouwman}, \& {Mo{\'o}r}}]{szulagyi12}
{Szul{\'a}gyi}, J., {Pascucci}, I., {{\'A}brah{\'a}m}, P., {et~al.} 2012, \apj,
  759, 47, \dodoi{10.1088/0004-637X/759/1/47}

\bibitem[{{Thanathibodee} {et~al.}(2023){Thanathibodee}, {Molina}, {Serna},
  {Calvet}, {Hern{\'a}ndez}, {Muzerolle}, \&
  {Franco-Hern{\'a}ndez}}]{thanathibodee23}
{Thanathibodee}, T., {Molina}, B., {Serna}, J., {et~al.} 2023, \apj, 944, 90,
  \dodoi{10.3847/1538-4357/acac84}

\bibitem[{{Tody}(1993)}]{tody93}
{Tody}, D. 1993, in Astronomical Society of the Pacific Conference Series,
  Vol.~52, Astronomical Data Analysis Software and Systems II, ed. R.~J.
  {Hanisch}, R.~J.~V. {Brissenden}, \& J.~{Barnes}, 173

\bibitem[{{Tokovinin} {et~al.}(2013){Tokovinin}, {Fischer}, {Bonati},
  {Giguere}, {Moore}, {Schwab}, {Spronck}, \& {Szymkowiak}}]{tokovinin13}
{Tokovinin}, A., {Fischer}, D.~A., {Bonati}, M., {et~al.} 2013, \pasp, 125,
  1336, \dodoi{10.1086/674012}

\bibitem[{{Vernet} {et~al.}(2011){Vernet}, {Dekker}, {D'Odorico}, {Kaper},
  {Kjaergaard}, {Hammer}, {Randich}, {Zerbi}, {Groot}, {Hjorth}, {Guinouard},
  {Navarro}, {Adolfse}, {Albers}, {Amans}, {Andersen}, {Andersen}, {Binetruy},
  {Bristow}, {Castillo}, {Chemla}, {Christensen}, {Conconi}, {Conzelmann},
  {Dam}, {de Caprio}, {de Ugarte Postigo}, {Delabre}, {di Marcantonio},
  {Downing}, {Elswijk}, {Finger}, {Fischer}, {Flores}, {Fran{\c{c}}ois},
  {Goldoni}, {Guglielmi}, {Haigron}, {Hanenburg}, {Hendriks}, {Horrobin},
  {Horville}, {Jessen}, {Kerber}, {Kern}, {Kiekebusch}, {Kleszcz}, {Klougart},
  {Kragt}, {Larsen}, {Lizon}, {Lucuix}, {Mainieri}, {Manuputy}, {Martayan},
  {Mason}, {Mazzoleni}, {Michaelsen}, {Modigliani}, {Moehler}, {M{\o}ller},
  {Norup S{\o}rensen}, {N{\o}rregaard}, {P{\'e}roux}, {Patat}, {Pena}, {Pragt},
  {Reinero}, {Rigal}, {Riva}, {Roelfsema}, {Royer}, {Sacco}, {Santin},
  {Schoenmaker}, {Spano}, {Sweers}, {Ter Horst}, {Tintori}, {Tromp}, {van
  Dael}, {van der Vliet}, {Venema}, {Vidali}, {Vinther}, {Vola}, {Winters},
  {Wistisen}, {Wulterkens}, \& {Zacchei}}]{vernet11}
{Vernet}, J., {Dekker}, H., {D'Odorico}, S., {et~al.} 2011, \aap, 536, A105,
  \dodoi{10.1051/0004-6361/201117752}

\bibitem[{{Wells} {et~al.}(2015){Wells}, {Pel}, {Glasse}, {Wright},
  {Aitink-Kroes}, {Azzollini}, {Beard}, {Brandl}, {Gallie}, {Geers}, {Glauser},
  {Hastings}, {Henning}, {Jager}, {Justtanont}, {Kruizinga}, {Lahuis}, {Lee},
  {Martinez-Delgado}, {Mart{\'\i}nez-Galarza}, {Meijers}, {Morrison},
  {M{\"u}ller}, {Nakos}, {O'Sullivan}, {Oudenhuysen}, {Parr-Burman}, {Pauwels},
  {Rohloff}, {Schmalzl}, {Sykes}, {Thelen}, {van Dishoeck}, {Vandenbussche},
  {Venema}, {Visser}, {Waters}, \& {Wright}}]{wells15}
{Wells}, M., {Pel}, J.~W., {Glasse}, A., {et~al.} 2015, \pasp, 127, 646,
  \dodoi{10.1086/682281}

\bibitem[{{Wright} {et~al.}(2023){Wright}, {Rieke}, {Glasse}, {Ressler},
  {Garc{\'\i}a Mar{\'\i}n}, {Aguilar}, {Alberts}, {{\'A}lvarez-M{\'a}rquez},
  {Argyriou}, {Banks}, {Baudoz}, {Boccaletti}, {Bouchet}, {Bouwman}, {Brandl},
  {Breda}, {Bright}, {Cale}, {Colina}, {Cossou}, {Coulais}, {Cracraft}, {De
  Meester}, {Dicken}, {Engesser}, {Etxaluze}, {Fox}, {Friedman}, {Fu},
  {Gasman}, {G{\'a}sp{\'a}r}, {Gastaud}, {Geers}, {Glauser}, {Gordon},
  {Greene}, {Greve}, {Grundy}, {G{\"u}del}, {Guillard}, {Haderlein},
  {Hashimoto}, {Henning}, {Hines}, {Holler}, {Detre}, {Jahromi}, {James},
  {Jones}, {Justtanont}, {Kavanagh}, {Kendrew}, {Klaassen}, {Krause},
  {Labiano}, {Lagage}, {Lambros}, {Larson}, {Law}, {Lee}, {Libralato}, {Lorenzo
  Alverez}, {Meixner}, {Morrison}, {Mueller}, {Murray}, {Mycroft}, {Myers},
  {Nayak}, {Naylor}, {Nickson}, {Noriega-Crespo}, {{\"O}stlin}, {O'Sullivan},
  {Ottens}, {Patapis}, {Penanen}, {Pietraszkiewicz}, {Ray}, {Regan},
  {Roteliuk}, {Royer}, {Samara-Ratna}, {Samuelson}, {Sargent}, {Scheithauer},
  {Schneider}, {Schreiber}, {Shaughnessy}, {Sheehan}, {Shivaei}, {Sloan},
  {Tamas}, {Teague}, {Temim}, {Tikkanen}, {Tustain}, {van Dishoeck},
  {Vandenbussche}, {Weilert}, {Whitehouse}, \& {Wolff}}]{wright23}
{Wright}, G.~S., {Rieke}, G.~H., {Glasse}, A., {et~al.} 2023, \pasp, 135,
  048003, \dodoi{10.1088/1538-3873/acbe66}

\end{thebibliography}
\end{document}